\documentstyle[12pt]{article}
\title{Is it always possible to discover supersymmetry broken at TeV scale 
at LHC?}
\author{\large N.V.~Krasnikov  \\[3mm]
\em Institute for Nuclear Research RAS, \\
\em Moscow, 117312, Russia  }
\begin{document}
\maketitle
\begin{abstract}
We show that the search for supersymmetry at LHC will be very 
problematic for the particular case of nonuniversal relations among gaugino 
masses. Namely, if   gluino, first chargino and LSP masses 
are closed to each other it would be very difficult to discover 
supersymmetry even if sparticle masses are lighter than 1 TeV.  
\end{abstract}

\newpage
Supersymmetric electroweak models offer the simplest solution of 
the gauge hierarchy problem \cite{1} -\cite{4}. In real life supersymmetry 
has to be broken, and the masses of superparticles must be lighter 
than $O(1)$ TeV \cite{4}.
The scientific program at the large hadron collider (LHC) \cite{5}-\cite{7} 
which will be the largest particle-accelerator complex ever built in the 
world has many goals. Among them  the discovery of the supersymmetry 
broken at TeV scale with sparticle masses less than $O(1)$ TeV  is the most 
important one. For the supersymmetric extension of the Weinberg-Salam 
model, soft supersymmetry breaking terms usually consist of the gaugino 
mass terms , squark and slepton masses and trilinear soft scalar 
terms. In general soft supersymmetry breaking terms are arbitrary. 
Within the minimal SUGRA-MSSM 
framework \cite{8} it would be possible to discover supersymmetry with 
squark and gluino masses up to (2 - 2.5) TeV \cite{9,10}. 
The standard signatures proposed for the search for squarks and gluino at LHC 
are \cite{5} -\cite{7}
\begin{equation}
jets + E^{T}_{miss},
\end{equation}

\begin{equation}
jets +  (n\geq 1) leptons + E^{T}_{miss}
\end{equation}
In SUGRA-MSSM framework all sparticle masses are determined mainly 
by two parameters: 
$m_0$(common squark and slepton mass at GUT scale)  and 
$m_{\frac{1}{2}}$(common gaugino mass at GUT scale). However, in general, 
due to many reasons we can expect that real sparticle masses can differ in a 
drastic way from sparticle masses pattern of SUGRA-MSSM model \cite{11}-
\cite{14}. Therefore, it is more appropriate to investigate LHC SUSY 
discovery potential in a model-independent way. Some prelimenary results 
in this direction have been obtained in refs. \cite{15,16}.
In particular it is 
very important to answer the question: is it always possible to 
discover supersymmetry broken at TeV scale at LHC for the case of arbitrary 
sparticle masses. 

In this paper we show that the search for supersymmetry at LHC will be 
very problematic for the particular case of nonuniversal relations 
among gaugino masses. Namely, for the case when gluino, 
first chargino and LSP 
masses are closed to each other it would be very difficult or even 
impossible to discover supersymmetry at LHC even if sparticle masses are 
lighter than 1 TeV. We assume that R-parity is conserved.

To be concrete consider the case when gluino, first chargino, second 
neutralino, LSP(lightest stable particle $\tilde{\chi}^0_1$) , squark 
and slepton masses are $m_{\tilde{g}}=500$ GeV, 
$m_{\tilde{\chi}^{\pm}_1}=m_{\tilde{\chi}^0_2} = 480$ GeV, 
$m_{\tilde{\chi}^0_1} = 450$ GeV, $m_{\tilde{q}} = m_{\tilde{l}} = 
600$ GeV. For such sparticle masses the search for direct slepton 
pair and gaugino $\tilde{\chi}^{\pm}_1 \tilde{\chi^0_2} $ productions 
is hopeless at LHC due to small cross sections. So we can expect to 
detect only strongly interacting particles(squarks, gluino) production 
using signatures (1,2). Consider gluino pair production
$pp \rightarrow \tilde{g} \tilde{g} + ...$ . Gluino decays 
$\tilde{g} \rightarrow \bar{q}q\tilde{\chi}^0_2$ and 
$\tilde{g} \rightarrow \bar{q}q^{'}\tilde{\chi}^{\pm}_1$ are suppresed in 
comparison with gluino decay into quark-antiquark pair and LSP 
$\tilde{g} \rightarrow \bar{q}q\tilde{\chi}^0_1$.  Hence the signature (2) 
which arises as a result of leptonic decays 
$\tilde{\chi}^0_2 \rightarrow l^+l^- \tilde{\chi}^0_1$ and 
$\tilde{\chi}^{\pm}_1 \rightarrow l^{\pm}\nu \tilde{\chi}^0_1$ is useless 
for the search for supersymmetry at LHC. The gluino decay mode 
$\tilde{g}\rightarrow \bar{q}q \tilde{\chi}^0_1$ leads to the signature (1). 
However for such values of gluino and LSP masses LSP particle is soft in 
gluino centre of mass frame. In parton model gluino are pair produced with 
small total value of transverse momentum $p_T$,  therefore in our case 
the average missing transverse energy $E_{miss}^T$ is rather small 
and it is determined by the mass difference 
$m_{\tilde{g}} - m_{\tilde{\chi}^0_1} = 50$ GeV. For such small values of 
$E^T_{miss}$ SM background is much bigger than signal that prevents the use 
of the signature (1) for gluino detection. For the squark pair production 
$pp \rightarrow \tilde{q}\tilde{q}^{'}+ ...$ the main squark decay mode is 
$\tilde{q} \rightarrow \tilde{g} q $   with soft gluino. Again in this case 
the signature (2) is not useful. For the signature (1) the typical 
$E^T_{miss}$ is less than 100 GeV that prevents SUSY discovery due to 
huge SM background. 

We have made simulations   at the particle level 
with parametrised detector responses based on a detailed detector 
simulation. We have made our concrete calculations for CMS detector \cite{5}. 
The CMS detector
simulation program CMSJET~3.2~\cite{17} has been  used. It incorporates 
the full electro-magnetic(ECAL)  and hadronic (HCAL) calorimeter granularity, 
and includes main calorimeter system cracks in rapidity and azimuth. The 
energy resolutions for muons, electrons(photons), hadrons and jets are 
parametrised. Transverse and longitudinal shower profiles are also 
included through appropriate parametrisations.
All SUSY processes have been generated with ISAJET7.32, ISASUSY \cite{18}
In our paper we have used the results of the background simulations 
of refs.~\cite{5,19}. The main results of our 
simulations is that SM background dominates for both the signatures (1) and 
(2) and prevents SUSY observation.

For the second  example with $m_{\tilde{g}} = 800$ GeV, 
$m_{\tilde{\chi}^0_2} = 
m_{\tilde{\chi}^{\pm}_1} = 690$ GeV, $m_{\tilde{\chi}^0_1} = 650$ GeV,  
$m_{\tilde{q}} = m_{\tilde{l}} = 700$ GeV the main gluino and 
squark decay modes 
are $\tilde{g} \rightarrow \bar{q}\tilde{q}$, 
$\tilde{q} \rightarrow q \tilde{\chi}^0_1$. Again in this case for signatures 
(1,2) SM background dominates.

For the third  example with $m_{\tilde{g}} = 700$ GeV, $m_{\tilde{\chi}^0_2} = 
m_{\tilde{\chi}^{\pm}_1} = 750$ GeV, $m_{\tilde{\chi}^0_1} = 650$ GeV 
$m_{\tilde{q}} = m_{\tilde{l}} = 670$ GeV the decays of squarks and gluino 
into the first chargino and second neutralino are  prohibited by kinematics 
and the main gluino and squark modes are 
$\tilde{g} \rightarrow \bar{q}\tilde{q}$, 
$\tilde{q} \rightarrow q \tilde{\chi}^0_1$. Again in this case for the 
signature (1) SM background dominates.

Let us state the main results of this paper: 
standard signatures (1,2) used for the search for 
supersymmetry at LHC  not always  allow to discover supersymmetry at LHC 
even if sparticle masses are lighter than 1 TeV. Namely, the search for 
supersymmetry will be very problematic for the particular case 
when gluino, first chargino and LSP masses are closed to each 
other. Probably $e^+e^-$ Next Linear Coillider with total 
energy $E_{cm} = 2$ TeV will have better perspectives to discover 
supersymmetry with such sparticle masses by the measurement of cross section 
of $e^+e^-$ annihilation into hadrons.  
                                             
I am   indebted to the collaborators of INR 
Theoretical Division  for useful discussions and comments.

\end{document}